\documentclass[
twocolumn,
prc,
showpacs,
a4paper,
nofootinbib,
superscriptaddress,
floatfix,
preprintnumbers,
flushbottom,
aps
]{revtex4-1}

\usepackage{url}
\usepackage[%
        hyperindex,breaklinks,
	pdfstartview=FitH,
  pdfpagemode=UseNone
	]{hyperref}
\usepackage{breakurl}

\usepackage[
   centertags, % (default) center tags vertically
   sumlimits,  %(default) Place the subscripts and superscripts of summation
   intlimits,  % Like sumlimits, but for integral symbols.
   namelimits, % (default) Like sumlimits, but for certain 'operator names' such as
]{amsmath} %
\usepackage{amssymb}

\makeatletter
\def\tagform@#1{\maketag@@@{\ignorespaces#1\unskip\@@italiccorr}}
\let\orgtheequation\theequation
\def\theequation{(\orgtheequation)}
\makeatother

\usepackage[all]{hypcap} % Links to figures point actually on figures itself, not on caption
\usepackage{dcolumn}

\usepackage[english]{babel} 

\usepackage[%
]{graphicx}

\usepackage{color}
\usepackage[fixpdftex,usenames,dvipsnames]{xcolor}
\definecolor{dark-gray}{gray}{0.3}

\usepackage{paralist}
\usepackage{slashed}

\newcommand{\breakeq}{\nonumber\\}
\newcommand{\arr}{\rightarrow}
\newcommand{\sqrs}{\sqrt{s}}
\def\l{\left}
\def\r{\right}
\def\d{{\rm d}}

\begin{document}
% \title{Radiative parton processes in perturbative QCD: the Gunion and Bertsch cross section compared to the exact result}
\title{Radiative parton processes in perturbative QCD - an improved version of the Gunion and Bertsch cross section from comparisons to the exact result}

\author{Oliver Fochler}
\email[E-mail: ]{fochler@th.physik.uni-frankfurt.de}

\author{Jan Uphoff}
\email[E-mail: ]{uphoff@th.physik.uni-frankfurt.de}
\affiliation{Institut f\"ur Theoretische Physik, Goethe-Universit\"at Frankfurt, Max-von-Laue-Str. 1, 
D-60438 Frankfurt am Main, Germany}

\author{Zhe Xu}
\affiliation{Department of Physics, Tsinghua University, Beijing 100084, China}

\author{Carsten Greiner}
\affiliation{Institut f\"ur Theoretische Physik, Goethe-Universit\"at Frankfurt, Max-von-Laue-Str. 1, 
D-60438 Frankfurt am Main, Germany}

\date{\today}

\begin{abstract}
In this work we compare the Gunion-Bertsch approximation of the leading order perturbative QCD radiation matrix element to the exact result. To this end, we revisit the derivation of the Gunion-Bertsch approximation as well as perform extensive numerical comparisons of the Gunion-Bertsch and the exact result. We find that when employing the matrix elements to obtain rates or cross sections from phase space integration, the amplitude by \textsc{Gunion} and \textsc{Bertsch} deviates from the correct result in characteristic regions of the phase space. We propose an improved version of the Gunion-Bertsch matrix element which agrees very well with the exact result in all phase space regions.
\end{abstract}

%\pacs{25.75.-q, 25.75.Bh, 25.75.Cj, 12.38.Mh, 24.10.Lx}

\maketitle

\section{Introduction}

When studying the dynamics of the partonic stage of relativistic heavy ion collisions based on microscopic perturbative QCD (pQCD) processes or on rate equations for such processes, it has been found that elastic interactions alone are not sufficient to describe the evolution of the system \cite{Geiger:1992si,Geiger:1992ac,Serreau:2001xq,Molnar:2001ux}. Due to the numerical complexity of such dynamic approaches the incorporation of inelastic particle production and annihilation processes has mostly been based on leading order matrix elements. A commonly used approximation to the leading order pQCD matrix element for partonic $2 \leftrightarrow 3$ processes is a result derived by \textsc{Gunion} and \textsc{Bertsch} (GB) in 1981 \cite{Gunion:1981qs}. This approximation gives a comparatively simple expression for the gluon radiation amplitude in terms of the transverse momentum of the radiated gluon $k_{\perp}$ and the transverse exchanged momentum $q_{\perp}$. It has been widely used when solving transport problems \cite{Xiong:1992cu,Biro:1993qt,Srivastava:1996qd,Mustafa:1997pm,Wong:1996ta,Chen:2009sm,Chen:2010xk,Gossiaux:2010yx}, for example via rate equations or via microscopic transport approaches, most notably the microscopic transport model \emph{Boltzmann Approach to MultiParton Scatterings} (BAMPS) \cite{Xu:2004mz,Xu:2007aa,Xu:2008av,Fochler:2008ts}.
In Ref.~\cite{Das:2010hs,Abir:2010kc,Bhattacharyya:2011vy} some efforts are made to go beyond the soft GB approximation while still obtaining a relatively compact form.

Recently, the results obtained within the BAMPS framework, especially with respect to the computed shear viscosity \cite{Xu:2007ns}, have been challenged in a paper by \textsc{Chen} et al.\ \cite{Chen:2011km}, who claim that a mis- or double-counting of symmetry factors when applying the GB matrix element to inelastic processes might lead to an overestimation of interaction rates in BAMPS by a factor of 6. Also a recent work by \textsc{Zhang} \cite{Zhang:2012vi} has addressed this issue and finds numerical discrepancies between cross sections based on the GB approximation compared to full leading order results for the $2 \rightarrow 3$ amplitudes, which have also been known since the early 1980s \cite{Berends:1981rb, Ellis:1985er}. The numerical discrepancy found in Ref.~\cite{Zhang:2012vi} is much less pronounced than the one claimed in Ref.~\cite{Chen:2011km}. However, it is not fully clear how the results of these two works compare since both use different screening schemes.

In this work we will address these issues in great detail by providing extensive numerical comparisons between the GB approximation and the exact leading order pQCD result as well as by analytically revisiting the derivation of the GB approximation. We will show that the approximation by \textsc{Gunion} and \textsc{Bertsch} needs to be carefully corrected in certain phase space regions when being employed to the computation of cross sections or rates from the matrix element. To this end we propose an improved version of the GB matrix element which is valid in all regions of phase space. However, we want to emphasize that the deviations of rates computed in the original GB approximation compared to rates computed from the improved GB and the exact matrix element are caused deep within the approximations made by \textsc{Gunion} and \textsc{Bertsch} and are not given by simple symmetry factors as argued in Ref.~\cite{Chen:2011km}. 

The paper is organized as follows. After a short revisit of the GB calculation we propose an improved version of the matrix element in Sec.~\ref{sec:GB}. In Sec.~\ref{sec:exact_me} the exact matrix elements for $2\rightarrow 3$ processes are introduced, whereas the phase space integration needed to obtain a total cross section is outlined in Sec.~\ref{sec:total_cs}. Finally, we will compare the original and improved GB cross section with the exact result and discuss deviations.

\section{Gunion-Bertsch matrix element}
\label{sec:GB}

In this section we will argue that the approximation to the leading order pQCD matrix element for $qq' \arr qq'g$ processes (and more generally for other gluon radiation processes as well) as computed by \textsc{Gunion} and \textsc{Bertsch} \cite{Gunion:1981qs} needs to be carefully corrected when being applied to the computation of cross sections or rates. This correction consists of two parts, 
\begin{inparaenum}[\itshape a\upshape)]
\item keeping a kinematic factor $(1-x)^{2}$, where $x$ is the fraction of light cone momentum carried by the radiated gluon; and
\item respecting the symmetry of the process by explicitly combining results from $A^{+}=0$ and $A^{-}=0$ calculations, restricting the emission of gluons to the respective forward direction.
\end{inparaenum}

\begin{figure}
	\centering
\begin{minipage}[t]{0.49\linewidth}
\centering
\includegraphics[width=0.7\textwidth]{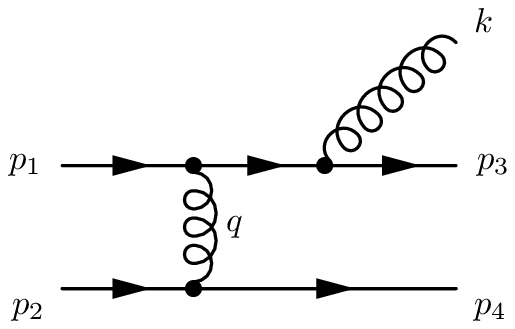}%eps

(1)

\vspace{1em}

\includegraphics[width=0.7\textwidth]{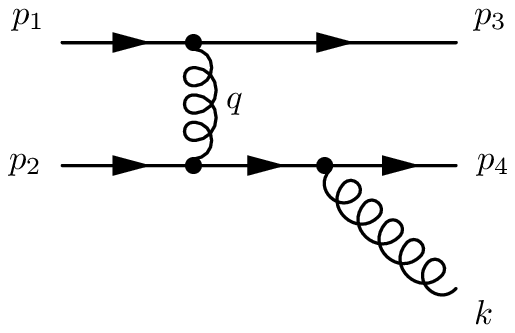}%eps

(3)
\end{minipage}
\hfill
\begin{minipage}[t]{0.49\linewidth}
\centering
\includegraphics[width=0.7\textwidth]{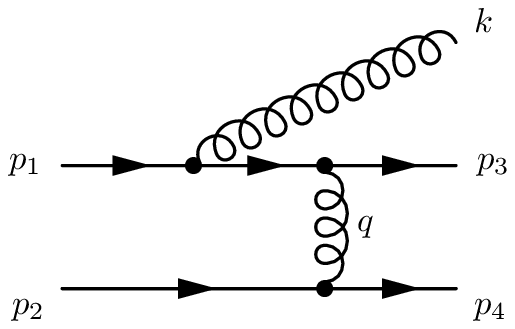}%eps

(2)

\vspace{1em}

\includegraphics[width=0.7\textwidth]{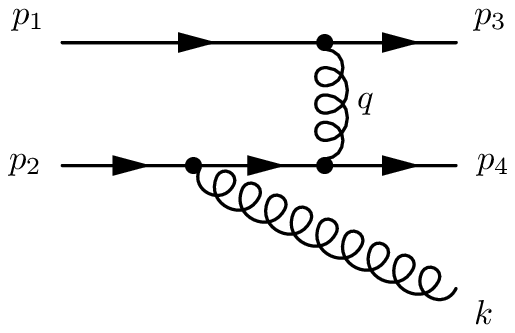}%eps

(4)
\end{minipage}

\vspace{1em}

\includegraphics[width=0.35\linewidth]{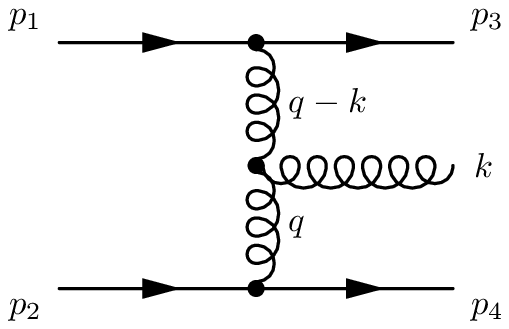}%eps

(5)
\caption{Feynman diagrams for $q+q' \arr q+q'+g$.}
\label{fig:feynman_diagrams_qq_qqg}
\end{figure}

We start by reviewing the derivation of the Gunion-Bertsch approximation, taking the example of $qq' \arr qq'g$ processes as in the original publication \cite{Gunion:1981qs}. In leading order this process is given by the five Feynman diagrams shown in \autoref{fig:feynman_diagrams_qq_qqg}.  

In the following we label the 4-momentum of the incoming (outgoing) quarks with $p_1$ and $p_2$ ($p_3$ and $p_4$). The radiated gluon is denoted with $k$, while $q$ is the internal momentum transfer (the momentum of the internal gluon propagator). Furthermore we are working in light cone coordinates where in the center of momentum frame the momenta of the incoming particles are given by
\begin{align}
	p_1^\mu &= \l( \sqrt{s}, 0, 0, 0 \r) \\
	p_2^\mu &= \l( 0, \sqrt{s}, 0, 0 \r) \ .
\end{align}
The Lorentz-invariant quantity
\begin{align}
x = k^+/p_1^+
\end{align}
characterizes the fraction of light cone momentum carried away by the radiated gluon. It can be related to the rapidity of the emitted gluon via
\begin{align}
\label{x_def_y}
x = \frac{k_\perp}{\sqrs} {\rm e}^y\ .
\end{align}
The \textsc{Gunion} and \textsc{Bertsch} matrix element is derived in the high energy limit, i.e.\ the radiated gluon and the momentum transfer of the process are soft,
\begin{align}
\label{gb_constrains_1}
	k_\perp &\ll \sqrt{s} \breakeq
	q_\perp &\ll \sqrt{s} \ .
\end{align}
The third approximation explicitly stated by GB relates $k_\perp$, $q_\perp$, and $x$,
\begin{align}
\label{gb_constrains_2}
	x q_\perp &\ll k_\perp \ .
\end{align}
With these approximations and from the constraint that all external particles are on-shell one can explicitly specify $k$ and $q$ in light cone coordinates
\begin{align}
	k^\mu &= \l( x \sqrs, \frac{k_\perp^2}{x\sqrs}, {\bf k}_\perp  \r) \label{eq:k_definition}\\
	q^\mu &= \l( -  \frac{q^2_\perp}{\sqrs}, \frac{k_\perp^{2}/x+q_\perp^{2}-2{\bf q}_{\perp}{\bf k}_{\perp}}{(1-x)\sqrs}, {\bf q}_\perp  \r) \ .
\end{align}

The original computation of the GB approximation has been carried out in light cone gauge, $A^+ = 0$. This implies that the $+$~component of the polarization vector $\epsilon(k)$ is also zero. The physical polarizations of the emitted gluon must be transverse to its momentum, $\epsilon(k) \cdot k = 0$. With these two constraints the two physical polarization vectors, $i=1,2$, are given by
\begin{align}
	\epsilon^\mu_{(i)}(k) = \l( 0, \frac{2 {\boldsymbol \epsilon}^{(i)}_\perp \cdot {\bf k}_\perp}{x \sqrs}, {\boldsymbol \epsilon}^{(i)}_\perp \r) \ ,
\end{align}
where ${\boldsymbol \epsilon}^{(1)}_\perp = (1,0)$ and ${\boldsymbol \epsilon}^{(2)}_\perp = (0,1)$ are possible choices. For brevity we will suppress the polarization index $(i)$ in the following, keeping in mind that the final polarization sum will simply amount to the replacement $\sum_{\epsilon} \left| {\bf p}\cdot{\boldsymbol \epsilon}_{\perp} \right|^{2} = {\bf p}^{2}$.

The evaluation of the Feynman diagrams in the GB approximation is detailed in \hyperref[sec:app_GB]{appendix~\ref*{sec:app_GB}}, we list the results here:

\begin{align}
\label{gb_me_single_diagrams}
i \mathcal{M}_1^{qq'\arr qq'g} &\simeq i \, g^{3} \, \lambda^a_{ij} \lambda^a_{mn} \lambda^b_{jl} \delta_{ws} \delta_{w's'} \frac{2s}{q_{\perp}^{2}} \, (1-x) \frac{2 \, {\bf k}_{\perp}\cdot {\boldsymbol \epsilon}_\perp}{k_{\perp}^{2}} \breakeq
i \mathcal{M}_2^{qq'\arr qq'g} &\simeq -i \, g^{3} \, \lambda^a_{jl} \lambda^a_{mn} \lambda^b_{ij} \delta_{ws} \delta_{w's'} \frac{2s}{q_{\perp}^{2}} \, (1-x)
\breakeq &\qquad \times \,  \frac{2 \, {\bf k}_{\perp}\cdot {\boldsymbol \epsilon}_\perp}{k_{\perp}^{2}} \breakeq
i \mathcal{M}_3^{qq'\arr qq'g} &\simeq 0 \breakeq
i \mathcal{M}_4^{qq'\arr qq'g} &= 0 \breakeq
i \mathcal{M}_5^{qq'\arr qq'g} &\simeq g^{3} \, f^{abc} \lambda^{a}_{il} \lambda^{c}_{mn} \delta_{ws} \delta_{w's'} \frac{2s}{q_{\perp}^{2}} \,  (1-x) \breakeq &\qquad \times \, \frac{2 \,({\bf q}_{\perp} - {\bf k}_{\perp})\cdot {\boldsymbol \epsilon}_\perp}{({\bf q}_{\perp} - {\bf k}_{\perp})^{2}}
\end{align}
Note that the diagrams corresponding to radiation from the lower quark line are either exactly (diagram 4) or approximately (diagram 3) zero in the $A^{+}=0$ gauge. The common $\frac{2s}{q_{\perp}^{2}}$ factors are already indicative of what after squaring the amplitude will be the factorized contribution of the elastic $qq' \arr qq'$ process in small angle approximation.

Using $[\lambda^a,\lambda^b]_{il} = i f^{abc} \lambda^c_{il}$ the sum of the five diagrams, Eq.~\eqref{gb_me_single_diagrams}, $i \mathcal{M}_{qq' \arr qq'g} = \sum_{j=1}^{5} i \mathcal{M}_j^{qq'\arr qq'g}$, can be computed
\begin{align}
\label{gb_me_all_diagrams}
	i \mathcal{M}_{qq' \arr qq'g} &\simeq - g^{3} f^{abc} \lambda^a_{mn} \lambda^c_{il} \delta_{ws} \delta_{w's'} \, \frac{2s}{q_{\perp}^{2}} \, (1-x) \breakeq 
		&\qquad \times 2 \, {\boldsymbol \epsilon}_\perp \cdot \l[ \frac{ {\bf k}_\perp}{k_\perp^2} +  \frac{ {\bf q}_\perp - {\bf k}_\perp}{({\bf q}_\perp - {\bf k}_\perp)^2} \r] \ .
\end{align}
Squaring the amplitude and taking the averages over the initial and sums over the final spin and color states yields 
\begin{align}
\label{gb_qqg_matrix_element_1minusX}
	{\l|\overline{\mathcal{M}}_{qq' \arr qq'g}\r|}^2 &\simeq \frac{24}{9} g^6  \, \frac{4s^{2}}{q_{\perp}^{4}} \, (1-x)^2 \, \l[ \frac{ {\bf k}_\perp}{k_\perp^2} +  \frac{ {\bf q}_\perp - {\bf k}_\perp}{({\bf q}_\perp - {\bf k}_\perp)^2} \r]^2 \breakeq
&= \frac{24}{9} g^6  \, \frac{4s^{2}}{q_{\perp}^{4}} \, (1-x)^2 \, \frac{ q_{\perp}^2}{ k_{\perp}^{2} ( {\bf q}_\perp - {\bf k}_\perp )^2 } \ .
\end{align}
Note that the algebraic simplification employed when going from the first to the second line of \eqref{gb_qqg_matrix_element_1minusX} is strictly only valid when not screening the propagator terms $1/k_{\perp}^{2}$ and $1/({\bf q}_\perp - {\bf k}_\perp)^2$ with a Debye mass, cf.~\autoref{sec:numerics}. When introducing the squared amplitude for the elastic $qq'\arr qq'$ process in the usual small angle (sa) approximation,  $\l|\overline{\mathcal{M}}_{qq'\arr qq'}\r|^2_{\text{sa}} = \frac{2}{9} g^{4} \, 4s^{2} / q_{\perp}^{4}$, Eq.~\eqref{gb_qqg_matrix_element_1minusX} can be rewritten as
\begin{multline}
\label{gb_qqg_matrix_element_factorized}
	{\l|\overline{\mathcal{M}}_{qq' \arr qq'g}\r|}^2 \simeq 12 g^2 \, \l|\overline{\mathcal{M}}_{qq'\arr qq'}\r|^2_{\text{sa}}  (1-x)^2
 \\ \times \, \frac{ q_{\perp}^2}{ k_{\perp}^{2} ( {\bf q}_\perp - {\bf k}_\perp )^2 } \ ,
\end{multline}
explicitly demonstrating that in the high energy limit the matrix element of the $2\arr 3$ process factorizes into an elastic $2\arr 2$ part and a gluon emission amplitude.

The factor $(1-x)^{2}$ in Eqs.~\eqref{gb_qqg_matrix_element_1minusX} or \eqref{gb_qqg_matrix_element_factorized} leads to a sizable suppression of the amplitude at forward rapidity, where $x > k_{\perp}/\sqrs$, which is immediately evident from Eq.~\eqref{x_def_y}. At the maximal rapidity that is kinematically allowed for a given transverse momentum, $y_{\text{max}} = \ln(k_{\perp}/\sqrs)$, $x$ is $1$ and the factor $(1-x)^2$ completely suppresses the matrix element. While \textsc{Gunion} and \textsc{Bertsch} have the $(1-x)$ terms in their intermediate results for the single diagrams, cf.\ \eqref{gb_me_single_diagrams}, they were mostly interested in emission spectra in the mid-rapidity region, $x \simeq k_{\perp} / \sqrs \ll 1$, and thus omitted the $(1-x)$ terms from their final results. The amplitude then simplifies to 
\begin{align}
\label{gb_qqg_matrix_element_org}
	{\l|\overline{\mathcal{M}}_{qq' \arr qq'g}\r|}^2 &\simeq 12 g^2  \,
	\l|\overline{\mathcal{M}}_{qq'\arr qq'}\r|^2_{\text{sa}}
	\frac{ q_{\perp}^2}{ k_{\perp}^{2} ( {\bf q}_\perp - {\bf k}_\perp )^2 } \ ,
\end{align}
which is the formula that has been mostly used in the literature as the GB matrix element since then \cite{Xiong:1992cu,Biro:1993qt,Srivastava:1996qd,Wong:1996ta,Mustafa:1997pm,Chen:2009sm,Chen:2010xk} and that has also been implemented in the transport model BAMPS \cite{Xu:2004mz}. When computing cross sections or rates from Eqs.~\eqref{gb_qqg_matrix_element_1minusX} or \eqref{gb_qqg_matrix_element_factorized}, however, the phase space integration covers the entire accessible rapidity region and factors $(1-x)$ must not be omitted. Thus Eqs.~\eqref{gb_qqg_matrix_element_1minusX} or \eqref{gb_qqg_matrix_element_factorized} rather than Eq.~\eqref{gb_qqg_matrix_element_org} need to be employed when computing rates for $2\arr 3$ processes in the GB approximation. 

The second modification that has been mentioned at the beginning of this section is somewhat more subtle. In order to arrive at the simple results for the single diagrams as noted in \eqref{gb_me_single_diagrams} when not setting $x \simeq k_{\perp} / \sqrs \simeq 0$, approximations are necessary that require (see \hyperref[sec:app_GB]{appendix~\ref*{sec:app_GB}} for more details)
\begin{equation} \label{eq:x2s_gg_kt_approximation}
 x^2 s \gg k_{\perp}^{2}
\end{equation}
or equivalently
\begin{equation} \label{eq:x2s_gg_kt_approximation_alternate_form}
 k^{+} \gg k^{-} \quad \Leftrightarrow \quad y \gg 0 \ .
\end{equation}
The results Eqs.~\eqref{gb_qqg_matrix_element_1minusX} and \eqref{gb_qqg_matrix_element_factorized} are thus only valid at mid-rapidity, where terms ${\sim} x$ can be neglected, and for forward emitted gluons, where $x^2 s \gg k_{\perp}^{2}$ holds, but not in the backward rapidity region. The requirement \eqref{eq:x2s_gg_kt_approximation} goes beyond the approximations \eqref{gb_constrains_1} and \eqref{gb_constrains_2} and has not been mentioned in the original GB publication.

Qualitatively, the need for a restriction such as Eqs.~\eqref{eq:x2s_gg_kt_approximation} or \eqref{eq:x2s_gg_kt_approximation_alternate_form} can be understood when noting that the process $qq' \arr qq'g$ is symmetric in interchanging $q$ and $q'$ and therefore the resulting amplitude should be symmetric in the rapidity of the emitted gluon. Terms including $x$, however, are evidently not symmetric in $y$, cf.\ Eq.~\eqref{x_def_y}, and thus the results \eqref{gb_qqg_matrix_element_1minusX} and \eqref{gb_qqg_matrix_element_factorized} do not obey the symmetry that should be present. This consideration nicely matches the analytic finding that \eqref{gb_qqg_matrix_element_1minusX} and \eqref{gb_qqg_matrix_element_factorized} are valid only in the forward and mid-rapidity region.

One can, however, carry out the same calculation in the $A^- = 0$ gauge. In this gauge diagrams 3, 4 and 5 have sizable contributions, while diagrams 1 and 2 do not contribute to the final amplitude. Setting
\begin{align}
\label{xdash_def_y}
x' = \frac{k_\perp}{\sqrs} {\rm e}^{-y}
\end{align} 
the final result for the averaged squared matrix element in $A^- = 0$ gauge reads 
\begin{multline}
\label{gb_qqg_matrix_element_xdash}
  {\l|\overline{\mathcal{M}}_{qq' \arr qq'g}\r|}^2 \simeq 12 g^2 \, \l|\overline{\mathcal{M}}_{qq'\arr qq'}\r|^2_{\text{sa}} (1-x')^2 \,
\\ \times \,  \frac{ q_{\perp}^2}{ k_{\perp}^{2} ( {\bf q}_\perp - {\bf k}_\perp )^2 } \ .
\end{multline}
This is simply the result obtained from the $A^+ = 0$ gauge, cf.~\eqref{gb_qqg_matrix_element_factorized}, with $x$ being replaced by $x'$. Since both results are also valid at mid-rapidity it is self-evident to combine Eqs.~\eqref{gb_qqg_matrix_element_1minusX} and \eqref{gb_qqg_matrix_element_xdash} to
\begin{align}
\label{gb_qqg_matrix_element_improved}
	{\l|\overline{\mathcal{M}}_{qq' \arr qq'g}\r|}^2 &\simeq 12 g^2  \,
	\l|\overline{\mathcal{M}}_{qq'\arr qq'}\r|^2_{\text{sa}} \,\frac{ q_{\perp}^2}{ k_{\perp}^{2} ( {\bf q}_\perp - {\bf k}_\perp )^2 }  
	%\l[ \frac{ {\bf k}_\perp}{k_\perp^2} +  \frac{ {\bf q}_\perp - {\bf k}_\perp}{({\bf q}_\perp - {\bf k}_\perp)^2} \r]^2 
	\breakeq
	&\quad \times \l[ (1-x)^2 \Theta(y) + (1-x')^2 \Theta(-y) \r]\breakeq 
		&= 12 g^2  \, \l|\overline{\mathcal{M}}_{qq'\arr qq'}\r|^2_{\text{sa}}  (1-\bar{x})^2 \breakeq &\quad \times \frac{ q_{\perp}^2}{ k_{\perp}^{2} ( {\bf q}_\perp - {\bf k}_\perp )^2 } \,
% \breakeq 
% 		&\qquad 
% 	\l[ \frac{ {\bf k}_\perp}{k_\perp^2} +  \frac{ {\bf q}_\perp - {\bf k}_\perp}{({\bf q}_\perp - {\bf k}_\perp)^2} \r]^2 \ ,
\end{align}
where we have defined
\begin{align}
\label{xbar_def_y}
\bar{x} = \frac{k_\perp}{\sqrs} {\rm e}^{|y|}\ .
\end{align}
In the following we will refer to this result as the improved GB matrix element since it is not only valid at mid-rapidity, but also at forward and backward rapidity. In Sec.~\ref{sec:numerics} we will compare this result to the exact matrix element without any approximation, which is briefly discussed in Sec.~\ref{sec:exact_me}. 

As a note, the idea of decomposing the phase space in forward and backward rapidities was also used in Ref.~\cite{guiho_phd} for the inelastic scattering of a light and heavy quark in the GB approximation, although the  $(1-\bar x)^2$ factor was neglected.

The observed factorization of the GB matrix element into an elastic part and a gluon emission amplitude in the high energy limit, cf.\ Eq.\eqref{gb_qqg_matrix_element_factorized}, immediately implies that the GB calculation is also valid for other $2 \arr 3$ processes, such as $qg \arr qgg$ or $gg \arr ggg$. In the high energy limit of the GB approximations the specific nature of the scattering particles becomes irrelevant for the structure of the resulting matrix elements. Therefore, Eq.~\ref{gb_qqg_matrix_element_improved} also holds for processes other than the one explicitly considered here, if one takes into account the different color prefactors of the corresponding $2 \arr 2$ small angle amplitudes, 
\begin{align} \label{eq:color_factors}
\l|\overline{\mathcal{M}}_{gg\arr gg}\r|^2_{\text{sa}} = \frac{9}{4} \, \l|\overline{\mathcal{M}}_{qg\arr qg}\r|^2_{\text{sa}} = \l(\frac{9}{4}\r)^2 \l|\overline{\mathcal{M}}_{qq'\arr qq'}\r|^2_{\text{sa}} \ .
\end{align}

\section{Exact matrix element}
\label{sec:exact_me}
The exact matrix elements for the $2 \arr 3$ processes involving light partons have been calculated in Refs.~\cite{Berends:1981rb, Ellis:1985er}. For $qq' \arr qq'g$ the result is relatively simple, \cite{Berends:1981rb}
\begin{widetext}
\begin{multline}
\label{exact_qqg_matrix_element}
{\l|\overline{\mathcal{M}}_{qq' \arr qq'g}\r|}^2 = \frac{g^6}{8} 
  \left[ \l( s^2 + s'^2 + u^2 + u'^2 \r) / t t' \right]  \left[ \l( p_1 k \r) \l( p_2 k \r) \l( p_3 k \r) \l( p_4 k \r) \right]^{-1} \\
  \times \left\{ C_1 \l[ \l( u + u' \r) \l( ss' + tt' - uu' \r) + u (st + s't') + u' (st' + s't )\r]  \r.\\
   \l. - C_2 \left[ \l( s + s' \r) \l( ss' - tt' - uu' \r) + 2 tt' (u+u') + 2uu' (t+t') \right] \right\} \ ,
\end{multline}
\end{widetext}
with the constants $C_1 = (N^2-1)^2/4N^3$ and $C_2 = (N^2-1)/4N^3$ where $N=3$ is the number of colors. The Mandelstam variables for the $2 \arr 3$ process are defined as
\begin{align}
 s &=(p_1+p_2)^2  & t&= (p_1-p_3)^2  & u&=(p_1-p_4)^2  \nonumber \\
 s^\prime &=(p_3+p_4)^2 & t^\prime&=(p_2-p_4)^2 & u^\prime&=(p_2-p_3)^2 
\end{align}
However, the expression for the matrix element has been simplified to such an extent that it is not obvious anymore how to identify the internal propagators which are usually screened with a mass of the order of the Debye mass in thermal QCD. We do not explicitly quote the result for $qg \arr qgg$ here, it is also given in \cite{Berends:1981rb}.

The expression for $gg \arr ggg$ is considerably more complicated, but can still be expressed in a relatively compact form due to its symmetry \cite{Berends:1981rb}
\begin{widetext}
\begin{multline} \label{eq:exact_me_gg_ggg}
{\l|\overline{\mathcal{M}}_{gg \arr ggg}\r|}^2 = \frac{g^6}{2} \left[ N^3 / (N^2 - 1) \right] \left[ (12345) + (12354) + (12435) + (12453) + (12534)\right.\\
 + \left. (12543) + (13245) + (13254) + (13425) + (13524) + (14235) + (14325) \right] \\
%  \times \left[ \frac{\left[ (p_{1} p_{2})^4 + (p_{1} p_{3})^4 + (p_{1} p_{4})^4 + (p_{1} p_{5})^4 + (p_{2} p_{3})^4 \right]}{  (p_{1} p_{2}) (p_{1} p_{3}) (p_{1} p_{4}) (p_{1} p_{5}) (p_{2} p_{3}) (p_{2} p_{4}) (p_{2} p_{5}) (p_{3} p_{4}) (p_{3} p_{5}) (p_{4} p_{5})  } \right.\\
%  + \left. \frac{ \left[(p_{2} p_{4})^4 + (p_{2} p_{5})^4 + (p_{3} p_{4})^4 + (p_{3} p_{5})^4 + (p_{4} p_{5})^4 \right]}{ (p_{1} p_{2}) (p_{1} p_{3}) (p_{1} p_{4}) (p_{1} p_{5}) (p_{2} p_{3}) (p_{2} p_{4}) (p_{2} p_{5}) (p_{3} p_{4}) (p_{3} p_{5}) (p_{4} p_{5}) } \right] \\
 \times \frac{(p_{1} p_{2})^4 + (p_{1} p_{3})^4 + (p_{1} p_{4})^4 + (p_{1} p_{5})^4 + (p_{2} p_{3})^4 + (p_{2} p_{4})^4 + (p_{2} p_{5})^4 + (p_{3} p_{4})^4 + (p_{3} p_{5})^4 + (p_{4} p_{5})^4 }{  (p_{1} p_{2}) (p_{1} p_{3}) (p_{1} p_{4}) (p_{1} p_{5}) (p_{2} p_{3}) (p_{2} p_{4}) (p_{2} p_{5}) (p_{3} p_{4}) (p_{3} p_{5}) (p_{4} p_{5})  }
\end{multline}
\end{widetext}
with $(ijklm) = (p_{i} p_{j}) (p_{j} p_{k}) (p_{k} p_{l}) (p_{l} p_{m}) (p_{m} p_{i})$ and $p_5$ being the momentum of the third outgoing gluon. In our notation $p_5=k$.
As for the other matrix elements, it is not a priori obvious how one would screen this matrix element. For the purpose of this work we introduce a cut-off prescription in Sec.~\ref{sec:numerics} that regulates the divergencies when calculating the total cross section and allows for consistent comparisons to the GB result.

\section{Phase space integration for cross sections}
\label{sec:total_cs}

In the previous sections the leading order pQCD matrix elements for radiative processes have been introduced both in their exact form and in the (improved) approximation by \textsc{Gunion} and \textsc{Bertsch}. These matrix elements depend on four independent phase space variables. A possible choice for these variables that is most appropriate for the GB approximation of the matrix element is shown later in this section, cf.\ Eq.~\eqref{total_cs_23_gb}. In the following we outline the procedure for the phase space integration that is needed to obtain the total cross section from the matrix elements presented in Sections \ref{sec:GB} and \ref{sec:exact_me}. Full energy and momentum conservation are taken into account. In general, differential cross sections (e.g. $\d \sigma / \d y$) can be obtained from this procedure by projecting out variables with delta functions (e.g. $\delta(y)$). It is these differential and total cross sections that are then used in \autoref{sec:numerics} to numerically check the improved GB approximation against the known exact result as they are easier to visualize than the bare matrix elements.

The total cross section of a $2 \rightarrow 3$ process is defined by
\begin{multline}
\label{total_cs_23}
 \sigma_{2 \rightarrow 3} = \frac{1}{2E_1 \, 2 E_2 \,  v_{\rm rel}} \frac{1}{\nu}
 \int \frac{\d^{3}p_{3}}{(2\pi)^{3}2E_{3}} \frac{\d^{3}p_{4}}{(2\pi)^{3}2E_{4}} \frac{\d^{3}k}{(2\pi)^{3}2k^0} \\
  \times\, (2\pi)^{4} \delta^{(4)}(p_{1}+p_{2} - (p_{3}+p_{4}+k) )
  \left| \overline{\mathcal{M}}_{2 \rightarrow 3} \right|^{2}
\ ,
\end{multline}
where we used the same notation as before. The relative velocity is defined by $v_{\rm rel}= \sqrt{(P_1^\mu P_{2 \mu})^2-m_1^2 m_2^2}/E_1 E_2$ and the symmetry factor $\nu = n!$ accounts for $n$ identical particles in the final state. Note that this symmetry factor is not needed when computing cross sections from the (improved) GB matrix element. In this case the specific choice and identification of the outgoing momenta, where $p_{5}=k$ is the momentum of the radiated gluon, selects a specific configuration and thus obviates the need for a symmetry factor $1/\nu$. This is taken into account for \emph{all} calculations involving the original or improved GB matrix elements in this paper and has also been correctly implemented in BAMPS from the beginning \cite{Xu:2004mz}.

Equation~\ref{total_cs_23} can be used to calculate the total cross section of the GB and exact matrix elements by performing the nine-dimensional integration. However, it is numerically more efficient to analytically integrate out the delta function to end up with only a five dimensional integration, which can be reduced to a four dimensional integration by performing one angle integration directly. After this and switching to the GB coordinates from Sec.~\ref{sec:GB} the total cross section reads \cite{Xu:2004mz}
\begin{multline}
\label{total_cs_23_gb}
 \sigma_{2 \rightarrow 3} = \frac{1}{256\, \pi^{4}} \frac{1}{\nu} \frac{1}{s}  \int_0^{s/4} \d q_{\perp}^{2}\,
\int_0^{s/4} \d k_{\perp}^{2}\, 
\int_{y_{\rm min}}^{y_{\rm max}} \d y \;
\int_{0}^{\pi}   \d\phi\:  \\
\times \left| \overline{\mathcal{M}}_{2 \rightarrow 3} \right|^{2}
   \sum \left(\left. \frac{\partial F}{\partial y_3} \right|_{F=0}\right)^{-1}
   \ ,
\end{multline}
where $y_3$ denotes the rapidity of particle 3 and $\phi$ the angle between ${\bf q}_\perp$ and ${\bf k}_\perp$. The available phase space limits the rapidity to $y_{\max / \min } = \pm \, {\rm arccosh} \l[ \sqrs/2 k_\perp \r]$.
Due to the transformation of the delta function, $\left(\partial F/\partial y_3\right)^{-1}$ must be summed over the roots of
\begin{align}
 F   &= (p_1+p_2-p_3-k)^2 \breakeq
 & =  s - 2\sqrt{s} \left( q_{\perp} \, \cosh y_{3} +  k_{\perp} \cosh y \right)
 + 2 q_\perp k_\perp \cos \phi
 \breakeq
& \qquad + 2 q_{\perp} \; k_{\perp} \left( \cosh y_{3} \cosh y -  \sinh y_{3} \sinh y \right) \ .
\end{align}

We have numerically checked that both Eqs.~\eqref{total_cs_23} and \eqref{total_cs_23_gb} give the same results for the GB and exact matrix elements, respectively.

\section{Numerical comparison}
\label{sec:numerics}

We have carried out extensive numerical comparisons between the exact matrix element and the matrix element calculated in the GB approximation. In order to obtain a picture as complete as possible we have investigated differential cross sections such as ${\rm d} \sigma / {\rm d} y$, ${\rm d} \sigma / {\rm d} x$, ${\rm d} \sigma / {\rm d} k_\perp {\rm d} q_\perp$ in addition to the total cross section. All calculations take into account the full kinematics according to Eqs.~\eqref{total_cs_23} or \eqref{total_cs_23_gb}, ensuring energy and momentum conservation. 

Both the exact and GB matrix elements are divergent for infrared and collinear configurations. In thermal quantum field theory these divergencies can be cured by loop resummations of the propagator, which leads to an extra term with the self energy in the propagator. Phenomenologically, this self energy can be  mimicked with a Debye mass $m_D$ that modifies the propagator terms and makes them infrared-safe,
\begin{equation}
\label{eq:debye_screening_prescription}
\frac{1}{q_\perp^2} \rightarrow \frac{1}{q_\perp^2 + m_D^2} \ .
\end{equation}

However, due to their compact notation it is not straightforward to identify the propagator terms in the exact matrix elements. Therefore, for all numerical calculations of total or differential cross sections from both the exact and the GB matrix elements we choose a simple cut-off procedure to cure the divergencies. The integrand is set to zero when the scalar product of any incoming or outgoing 4-momenta is smaller than a cut-off $\Lambda^2$, $p_i \cdot p_j < \Lambda^2$ with $i,j = 1...5$. Formally this is expressed by multiplying the integrand in Eq.~\eqref{total_cs_23} with $\Theta ( p_i \cdot p_j - \Lambda^2 )$, where $\Lambda^2$ is chosen to be proportional to the Debye mass, 
\begin{equation}
\label{eq:lambda_epsilon}
\Lambda^2 = \epsilon \, m_D^2 \ .
\end{equation}
It is immediately obvious from Eq.~\eqref{eq:exact_me_gg_ggg} that this prevents divergencies of the integrand. For propagator terms this scheme acts similar to the Debye screening, but the restrictions to the phase space potentially go beyond Debye screening only propagator terms. The resulting numerical values for the total or differential cross sections need thus not be the physically correct values and are consequently given in arbitrary units where necessary. Still this cut-off prescription allows for consistent and well-defined comparisons between the GB approximation and the exact leading order matrix elements which is the focus of this study. If not mentioned otherwise, we set $\epsilon = 0.001$ to reduce the screening effect as much as possible. In the last part of the section we will discuss the impact of different $\epsilon$ and compare this cut-off procedure to the standard Debye screening procedure for more physical scenarios.

The calculations in the remainder of the section are done for a temperature of $T=400 \, {\rm MeV}$. The coupling is set constant, $\alpha_s = 0.3$. If not stated otherwise, we use the average thermal value for massless particles for the squared center of mass energy, $s=18 T^2 \simeq 2.88 \, {\rm GeV}^{2}$, and determine $\Lambda$ from the usual gluon Debye mass for Boltzmann statistics, $m_{D}^2 = \frac{8 \alpha_s}{\pi} (N_c+n_f) \, T^2$. For $n_{f} = 3$ at the given temperature and coupling, the Debye mass is $m_{D}^2 \simeq 0.73 \, {\rm GeV}^2$.

\begin{figure}
\includegraphics[width=1.0\linewidth]{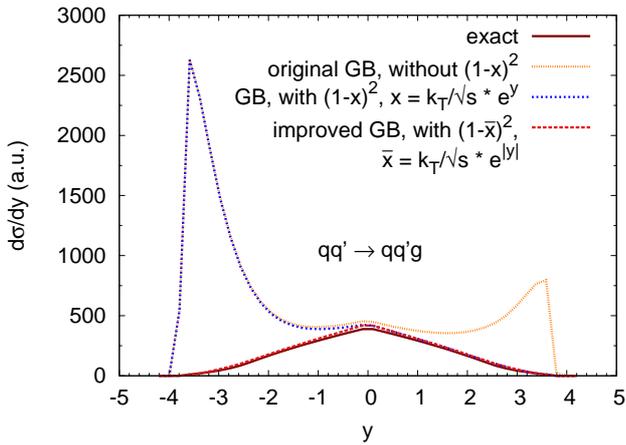}%eps
\caption{Differential cross section ${\rm d} \sigma / {\rm d} y$ for the process $qq' \rightarrow qq'g$ calculated with the exact (Eq.~\eqref{exact_qqg_matrix_element}), GB \eqref{gb_qqg_matrix_element_org}, GB with $(1-x)^2$ \eqref{gb_qqg_matrix_element_1minusX}, and improved GB with $(1-\bar{x})^2$ \eqref{gb_qqg_matrix_element_improved} matrix element.}
\label{fig:ds_dy_qq}
\end{figure}

Figure~\ref{fig:ds_dy_qq} compares the rapidity spectrum of the emitted gluon as given by different approximations of the matrix element by depicting the differential cross section ${\rm d} \sigma / {\rm d} y$ for the process $qq' \rightarrow qq'g$. %
This process is the clearest to study since only one (the emitted) gluon is involved. Furthermore, it is the process that has been studied in the original GB publication, cf.\ \autoref{sec:GB}. The plot nicely demonstrates the shortcomings of the standard GB matrix element, Eq.~\eqref{gb_qqg_matrix_element_org}, in the forward and backward region. 

The exact matrix element, Eq.~\eqref{exact_qqg_matrix_element}, only peaks at mid-rapidity and does not have any sizable contribution at forward or backward rapidity. The gluon emission into the forward and backward region is suppressed since the phase space is occupied by the two quarks which just scatter off with a small angle. Furthermore, the cross section is symmetric in the rapidity of the emitted gluon as it must be for this process. The standard GB matrix element without the $(1-x)^2$ term (Eq.~\eqref{gb_qqg_matrix_element_org}) is very similar at mid-rapidity, but has two additional large contributions at forward and backward rapidity. Thus this matrix element would allow that, for instance, the gluon is emitted into the backward region and that one of the quarks is located at mid-rapidity. Although the matrix element itself is symmetric in $y$, the curve is not since in our implementation the $p_z$ of particle $3$ is restricted to positive values to ensure that the particle only scatters off with a small angle.\footnote{This constraint is necessary if one transforms from Mandelstam $t$ to $q_\perp^2$ for such a small angle favoring process, since $q_\perp^2$ can be small for a outgoing particle which moves exactly in the other direction than it moved before the reaction, but in this case the scattering angle is not small anymore, thus, $t$ becomes large and $t \simeq q_\perp^2$ does not hold anymore.} %Obviously, to make the cross section symmetric again one also has to implement such a constraint for particle 4 as well, forbidding that it has a positive $p_z$.% 

The matrix element with the GB approximation and the term $(1-x)^2$, Eq.~\eqref{gb_qqg_matrix_element_1minusX}, has the same value at mid-rapidity as the other matrix elements since $x$ is small in this region. However, at forward rapidity the $(1-x)^2$ factor leads to a significant reduction compared to the pure GB result. Here the curve lies right on top of the curve from the exact matrix element. Nevertheless, as discussed in Sec.~\ref{sec:GB}, the curve with  $(1-x)^2$ is not symmetric anymore due to the asymmetric $y$-dependence contained in $x$. The last curve in Fig.~\ref{fig:ds_dy_qq} shows the cross section of our proposed improvement of the GB matrix element, Eq.~\eqref{gb_qqg_matrix_element_improved}. At forward and mid-rapidity it agrees with the previously mentioned curve. At backward rapidity the modified $(1-\bar{x})^2$ term removes the excess and reconciles the shape of the GB approximation with the symmetry of the process. 

\begin{figure}
\includegraphics[width=1.0\linewidth]{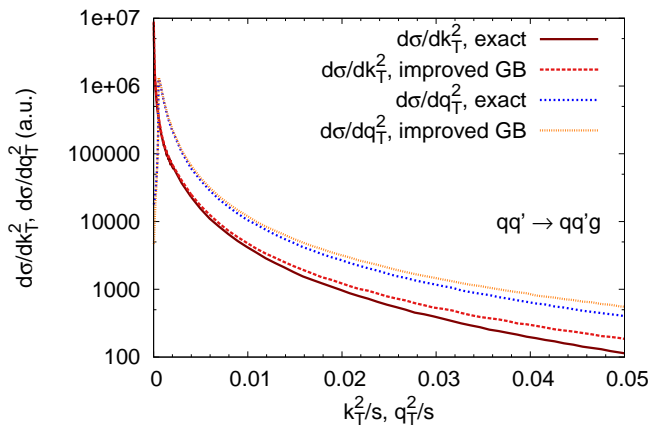}%eps
\caption{Differential cross sections ${\rm d} \sigma / {\rm d} k_\perp^2$ and ${\rm d} \sigma / {\rm d} q_\perp^2$ as a function of $k_T^2/s$ and $q_T^2/s$, respectively, for the process $qq' \rightarrow qq'g$ calculated with the exact (Eq.~\eqref{exact_qqg_matrix_element}) and improved GB (Eq.~\eqref{gb_qqg_matrix_element_improved}) matrix element.}
\label{fig:ds_dkt_dqt_qq}
\end{figure}

The overall agreement between the differential cross sections from the improved matrix element and from the exact result is remarkably good. The difference between the approximation and the exact result is on the level of a few percent over the entire $y$-range. This striking agreement is also visible in other differential cross sections such as ${\rm d} \sigma / {\rm d} q_\perp^2 {\rm d} k_\perp^2$ and ${\rm d} \sigma / {\rm d} x$. In Fig.~\ref{fig:ds_dkt_dqt_qq} the transverse momentum distributions ${\rm d} \sigma / {\rm d} k_\perp^2$ and ${\rm d} \sigma / {\rm d} q_\perp^2$ are depicted. 
At small $k_T^2$ and $q_T^2$, where the contribution to the cross section is largest, the deviation between the improved GB and the exact result is less than 5\,\%. The asymmetry between the $k_T^2$ and $q_T^2$ distributions is due to an interplay between the cut-offs and the $(1-x)$ factor for the GB matrix element. The presence of this asymmetry in the exact curves is a further indicator that the $(1-x)$ factor is essential in describing the exact result.

\begin{figure}
\includegraphics[width=1.0\linewidth]{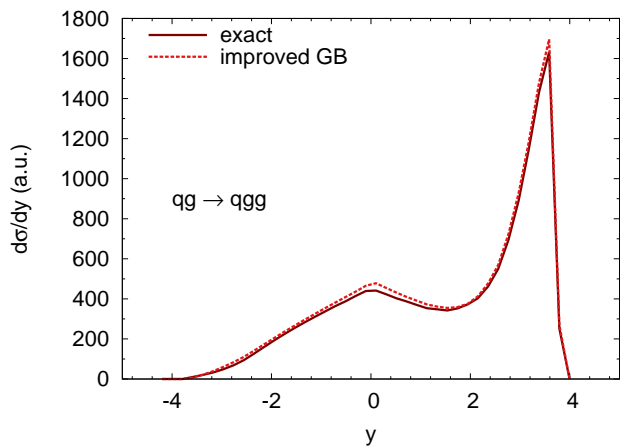}%eps
\caption{Differential cross section ${\rm d} \sigma / {\rm d} y$ for all outgoing gluons for the process $qg \rightarrow qgg$ calculated with the exact and improved GB matrix element.}
\label{fig:ds_dy_gq}
\end{figure}

Figure~\ref{fig:ds_dy_qq} was done for the simplest process, $qq' \rightarrow qq'g$, where only one gluon is involved. However, the findings also hold for $qg \rightarrow qgg$ and $gg \rightarrow ggg$. Since gluons are indistinguishable particles we plot the ${\rm d} \sigma / {\rm d} y$ for $qg \rightarrow qgg$ of all outgoing gluons in Figure~\ref{fig:ds_dy_gq}. 
Since we choose the incoming gluon to be particle $1$ and the process favors small angle scattering, there is a peak at forward rapidity in addition to the gluon peak at mid-rapidity. Again the curve from the improved GB matrix element reproduces the curve from the exact one very well. This also holds for $gg \rightarrow ggg$, where one gets an additional gluon peak in the backward region which corresponds to the third outgoing gluon.

\begin{figure}
\includegraphics[width=1.0\linewidth]{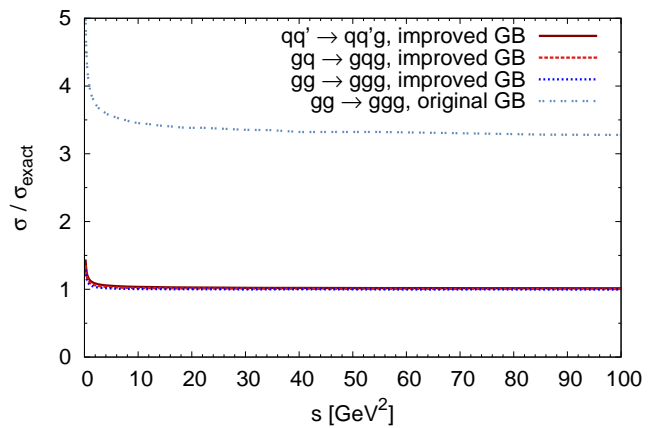}%eps
\caption{Ratio of total cross section $ \sigma_{2 \rightarrow 3} $  of the improved GB matrix element (Eq.~\eqref{gb_qqg_matrix_element_improved}) to the exact one (see Sec.~\ref{sec:exact_me}) as a function of $s$ for $qq' \rightarrow qq'g$, $qg \rightarrow qgg$, and $gg \rightarrow ggg$. In addition, the ratio of the original GB cross section \eqref{gb_qqg_matrix_element_org} as it was previously implemented in BAMPS to the exact cross section is plotted for $gg \rightarrow ggg$.}
\label{fig:sigma_s}
\end{figure}

Since all differential cross sections for the improved GB and exact matrix elements agree nicely, it is not a surprise that this is also true for the total cross sections. In Figure~\ref{fig:sigma_s} the ratios of the total cross sections $ \sigma_{2 \rightarrow 3} $ for different processes from the improved GB to those obtained from the exact matrix elements are depicted as a function of the squared center of mass energy $s$. 
The exact and improved cross sections agree very well for all processes and virtually all $s$. Only at very small $s$, below the thermal average $s=18 T^2 \simeq 2.88 \, {\rm GeV}^{2}$, there is a slight discrepancy up to about 40\,\%. The original GB cross section for $gg \rightarrow ggg$ as it was previously implemented in BAMPS is about a factor of three larger than the exact cross section. The same is true for the other processes (not plotted).

\begin{figure}
\includegraphics[width=1.0\linewidth]{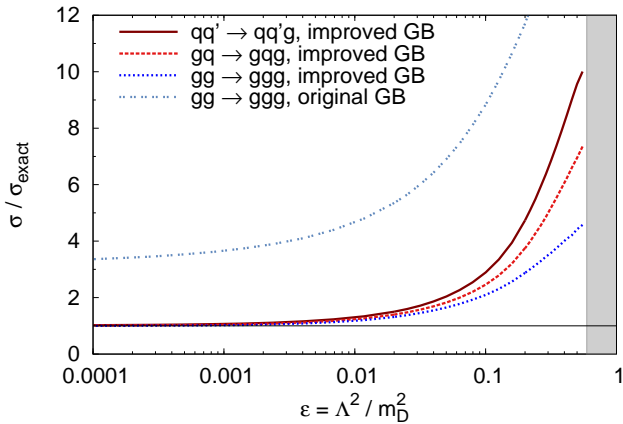}%eps
\caption{Ratio of the total cross section $ \sigma $ of the improved GB matrix element to the exact result  for $qq' \rightarrow qq'g$, $qg \rightarrow qgg$, and $gg \rightarrow ggg$ as a function of the screening cut-off $\epsilon$ from Eq.~\eqref{eq:lambda_epsilon}.  In addition, the ratio of the original GB cross section of $gg \rightarrow ggg$ to the exact is depicted as it was previously implemented in BAMPS.}
\label{fig:sigma_s_epsilon}
\end{figure}

Figure~\ref{fig:sigma_s_epsilon} shows the dependence of the total cross section on the cut-off $\epsilon = \Lambda^2 / m_D^2$.
The improved GB matrix element works best for small $\epsilon$ where the matrix elements are dominated by small $k_\perp$ and $q_\perp$. At larger $\epsilon$ most of the phase space at small $k_\perp$ and $q_\perp$ is cut away and other phase space regions play the dominant role, in which the GB approximation is not as good. Hence, the ratio rises. At $\epsilon \gtrsim 0.6$ (indicated by the grey band in the figure) the cut-off is so large that the entire available phase space is cut away and the GB as well as exact cross sections are zero. For such severe cut-offs the comparison scheme is not reliable any more and the previously introduced Debye screening would be more realistic. However, the comparison to the exact matrix element with standard Debye screening is not done in this paper since the identification of the propagators in the compact results of the exact matrix elements is not known. Also note that Fig.~\ref{fig:sigma_s_epsilon} is made for thermal $s=18T^2$. For larger values of $s$ the cut-off parameter $\epsilon$ at which the ratio deviates from unity becomes significantly larger. Thus, in summary, the improved GB approximation works best for larger $s$ and/or smaller screening cut-offs $\epsilon$.

\begin{figure}
\includegraphics[width=1.0\linewidth]{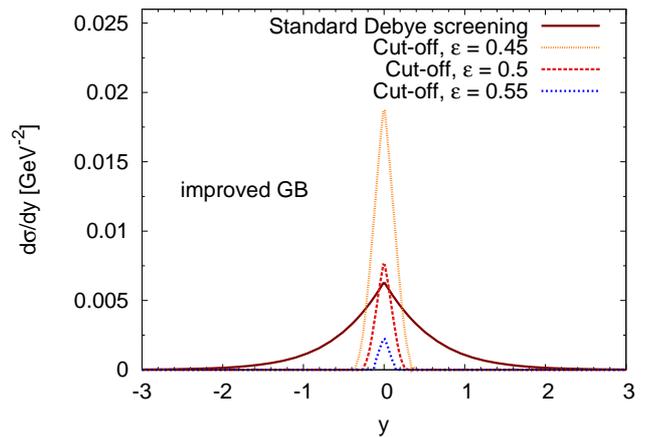}%eps
\caption{Differential cross section ${\rm d} \sigma / {\rm d} y$ for the process $qq' \rightarrow qq'g$ calculated with the improved GB matrix element with the standard Debye screening and the cut-off procedure for different $\epsilon$.}
\label{fig:dsigma_dy_epsilon_fit}
\end{figure}

Our improved GB result is therefore an especially good approximation for studying jet effects in heavy-ion collisions. For bulk dynamics, where the mean $s$ is smaller, it is important how strong the screening effects actually are and if one is already in the regime of large $\epsilon$ where the improved GB matrix element might deviate from the exact matrix element. To explore this in more detail, in Fig.~\ref{fig:dsigma_dy_epsilon_fit} we explicitly compare the improved GB differential cross section obtained from the cut-off procedure to the one obtained using the standard Debye screening according to Eq.~\eqref{eq:debye_screening_prescription}, which is for $qq' \arr qq'g$ (cf. Eqs.~\eqref{gb_qqg_matrix_element_1minusX} and \eqref{gb_qqg_matrix_element_improved})\footnote{The implementation in BAMPS does not screen the $1/k_\perp^2$ term in the bracket since small $k_\perp$ are regularised by the LPM effect~\cite{Fochler:2010wn}. However, since the LPM effect is not employed in this study, we regularise also this term with the Debye mass.}
\begin{multline}
	{\l|\overline{\mathcal{M}}_{qq' \arr qq'g}\r|}^2 =  
	 \frac{32}{3}  g^6	 \frac{s^2}{(q_\perp^2 + m_D^2)^2} 
	(1-\bar{x})^2 \\ 
	\times 
	\l[ \frac{ {\bf k}_\perp}{k_\perp^2 + m_D^2} +  \frac{ {\bf q}_\perp - {\bf k}_\perp}{({\bf q}_\perp - {\bf k}_\perp)^2 + m_D^2} \r]^2 \ .
\end{multline}
Now to get a feel for the quality of the improved GB approximation for Debye screened thermal processes one can either try to match the total cross sections or the differential cross sections ${\rm d} \sigma / {\rm d} y$ at mid-rapidity. Doing this, the necessary parameter $\epsilon$ can be determined to be of the order of $\epsilon \approx 0.4$ to $0.5$. As can be seen from Fig.~\ref{fig:sigma_s_epsilon} the deviations between the improved GB and exact result in this $\epsilon$ region vary roughly between $\sigma / \sigma_{\text{exact}} \approx 4$ and $\sigma / \sigma_{\text{exact}} \approx 9$. However,  Fig.~\ref{fig:dsigma_dy_epsilon_fit} also clearly demonstrates the distortion of the phase space due to extreme choices of the cut-off. Large rapidities are cut away entirely while the standard Debye screening procedure leaves the whole phase space available. It is questionable whether the strong cuts in the cut-off procedure make physical sense. The comparison is thus a rather qualitative one and should be regarded as a first estimate of the quality of the (improved) GB approximation in the region of thermal processes with a standard Debye screening.

\begin{figure}
\includegraphics[width=1.0\linewidth]{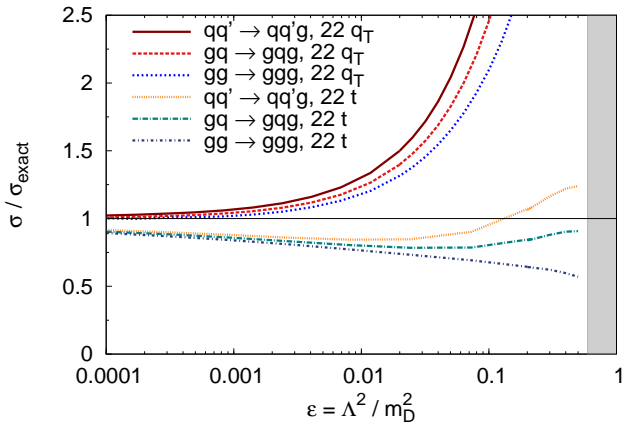}%eps
\caption{As Fig.~\ref{fig:sigma_s_epsilon} but in addition the improved GB cross section with the exact $2 \arr 2$ matrix element (22 $t$) instead of the $2 \arr 2$ matrix element in small angle approximation (22~$q_\perp$).}
\label{fig:sigma_s_epsilon_22t}
\end{figure}

As discussed in \autoref{sec:GB}, the factorization of the GB approximation into an elastic $2 \rightarrow 2$ part and a radiation amplitude features the elastic amplitude in small angle approximation, $\l|\overline{\mathcal{M}}_{2 \arr 2}\r|^2_{\text{sa}} \propto 1 / q_{\perp}^{4}$. If one, however, instead employs the exact binary matrix element for the $t$ channel which is
\begin{equation}
\label{eq:me_22_t}
\l|\overline{\mathcal{M}}_{2\arr 2}\r|^2 \propto \frac{1}{t^2}\ ,
\end{equation}
the agreement between the exact and the improved GB result becomes much better at large $\epsilon$, which is illustrated in Figure~\ref{fig:sigma_s_epsilon_22t}. 
The ratio is of the order of 1 over the entire $\epsilon$ range and the deviation is at most 50\,\% for all processes. In contrast to the previous result with the approximated $2\arr 2$ matrix element (cf. Fig.~\ref{fig:sigma_s_epsilon}), the ratio for small $\epsilon$ is not 1 anymore, but 0.9. This is probably due to the fact that different approximations are employed for the $2 \arr 2$ and the radiative part of the matrix element. However, since the numerical agreement is so good also in the region of thermal processes at larger $\epsilon$, the combination of the improved GB approximation~\eqref{gb_qqg_matrix_element_improved} for the radiation amplitude with the exact $2 \arr 2$ part~\eqref{eq:me_22_t} might be implemented in our transport model BAMPS.

\section{Summary}

The inelastic gluon Bremsstrahlung matrix element derived by \textsc{Gunion} and \textsc{Bertsch} for $qq' \arr qq'g$ has been compared to the exact matrix element for this process by performing full phase space integrations and analyzing the total and differential cross sections. Although it agrees very well with the exact result at mid-rapidity (the region in which \textsc{Gunion} and \text{Bertsch} were mainly interested in) large deviations have been found at forward and backward rapidity. Based on a detailed analytic investigation of the underlying approximations, we propose an improved version of the GB matrix element which agrees very well in all phase space regions with the exact result. This remarkable agreement holds for virtually all center of mass energies $\sqrt{s}$ and most cut-off parameters $\epsilon$ as well as for the other radiative processes $qg \arr qgg$ and $gg \arr ggg$.

In contrast to the claim in Ref.~\cite{Chen:2011km}, the deviation between the GB matrix element and the exact result does not originate from a miscounting of symmetry factors, but lies deep in the approximations of GB in forward and backward rapidities. Although the reasoning in Ref.~\cite{Chen:2011km} accidentally (and thus only effectively) holds for $gg \arr ggg$ due to the symmetry of the process, it fails for asymmetric processes such as $qg \arr qgg$ or $qq' \arr qq'g$.

The implementation of the new improved GB matrix element in our transport model BAMPS is currently under way. From Fig.~\ref{fig:sigma_s} we expect at most a factor of 3 difference in the rates, although this should be reduced by the iterative computation of cross sections due to the implementation of the LPM effect \cite{Fochler:2010wn}. The difference in the phase space sampling of the outgoing particles will have a large impact in particular on jet observables. For instance, we expect much less suppression of high energy particles, which will increase the small nuclear modification factor of BAMPS \cite{Fochler:2008ts,Fochler:2010wn} and bring it closer to the data. The effect on bulk observables will also be of great interest. Although we expect a decrease of the elliptic flow \cite{Xu:2008av} and an increase of the shear viscosity \cite{Xu:2007ns}, one must wait for the numerical results to see how large the effect will actually be. 
As part of this study we will also implement the running coupling for elastic and radiative processes which is expected to lead to a rising $R_{AA}$ curve with transverse momentum.
Furthermore, we plan to generalize the new improved GB matrix element also to the heavy flavor sector and study its impact on the heavy quark energy loss \cite{Abir:2011jb,Uphoff:2011ad,Uphoff:2012gb}.

\section*{Acknowledgements}

We would like to thank J.\ Aichelin, A.\ Dumitru, P.B.\ Gossiaux, and A.\ Peshier for stimulating and helpful discussions. This work was supported by the Bundesministerium f\"ur Bildung und Forschung (BMBF) and by the Helmholtz International Center for FAIR within the framework of the LOEWE program launched by the State of Hesse. Numerical computations have been performed at the Center for Scientific Computing (CSC).
J.U. is grateful for the kind hospitality at Tsinghua University, where part of this work has been done. Furthermore, J.U. would like to thank the DAAD and the Helmholtz Research School for Quark Matter studies for financial support.
Z.X. is supported by the NSFC under grant No.\ 11275103.

\appendix
\section{Detailed GB calculation} \label{sec:app_GB}

The purpose of this appendix is to provide some more details on the computation of the $qq' \arr qq'g$ matrix element in the \textsc{Gunion} and \textsc{Bertsch} approximation, cf.\ Sec.~\ref{sec:GB} and the diagrams shown in \autoref{fig:feynman_diagrams_qq_qqg}. The focus is on highlighting crucial steps and approximations that are needed to arrive at the results in Eq.~\eqref{gb_me_single_diagrams}.

We define
\begin{equation} \label{eq:def_Jrs}
 J_{rs}^{\mu}(p,p') = \overline{u}^{r}(p) \, \gamma^{\mu} \, u^{s}(p') \ ,
\end{equation}
where $s$ and $r$ are the spin indices of the spinors $u$ and $\overline{u}$. Setting $p'=p+q$ and assuming that $q \ll p$, the Gordon identity \cite{Peskin} for \eqref{eq:def_Jrs} can be simplified to give
\begin{equation}
J_{rs}^{\mu}(p,p+q) \simeq (2p+q)^{\mu} \, \delta_{rs} \ .
\end{equation}
This eikonal approximation essentially amounts to describing the problem in scalar QCD and is responsible for the fact that the amplitudes of different $2 \arr 3$ processes only differ by the color factor of the factorized elastic amplitude, cf.\ Eq.~\eqref{eq:color_factors}. Furthermore we define 
\begin{align}
 \Gamma_{rs}^{\mu\nu}(p,k,p') &= \overline{u}^{r}(p) \, \gamma^{\mu} \, ({\slashed k} + m) \, \gamma^{\nu} \, u^{s}(p') \breakeq
  &= \sum_{t} J_{rt}^{\mu}(p,k) J_{ts}^{\nu}(k,p') \ .
\end{align}

With these definitions and the kinematics as given in Sec.~\ref{sec:GB} we can compute the matrix elements of the diagrams.

\subsubsection*{Diagram 1}
The Feynman rules for diagram 1, cf.\ \autoref{fig:feynman_diagrams_qq_qqg}, give
\begin{align}
 i \mathcal{M}_1^{qq'\arr qq'g} &= \frac{F_{1}}{q^2 (p_{3}+k)^2} \, \Gamma_{ws}^{\alpha\mu}(p_{3},p_{3}+k,p_{1}) \breakeq &\qquad \times \epsilon_{\alpha}^{*}(k) \, g_{\mu\nu} \, J_{w's'}^{\nu}(p_{4},p_{2}) \breakeq
&\simeq \frac{F_{1}\,\delta_{ws} \delta_{w's'}}{q^2 (p_{1}+q)^2} \, (2p_1+q)^\mu (2p_2-q)_\mu \breakeq &\qquad \times (2p_1 + 2q - k)^\alpha \epsilon_{\alpha}^{*}(k)
\end{align}
with $F_{1} = -i \, g^{3} \, \lambda^a_{ij} \lambda^c_{mn} \lambda^b_{jl} \delta^{ac}$. Employing the kinematic approximations from Sec.\ \ref{sec:GB} to the single terms
\begin{align}
(p_{1}+q)^2 &\simeq \frac{({\bf k}_{\perp} - x{\bf q}_{\perp})^2}{(1-x)x}\\
(2p_1+q)^\mu (2p_2-q)_\mu &\simeq 2s \label{eq:M1_x2kt_approx}\\
(2p_1 + 2q - k)^\alpha \epsilon_{\alpha}^{*} &\simeq \frac{2}{x}({\bf k}_{\perp} - x{\bf q}_{\perp})\! \cdot\! {\boldsymbol \epsilon}_\perp
\end{align}
together with $q^2 \simeq -q_{\perp}^{2}$ and \eqref{gb_constrains_2} yields the result listed in Eq.~\eqref{gb_me_single_diagrams}. Eq.~\eqref{eq:M1_x2kt_approx} requires the restriction $x^2 s \gg k_{\perp}^{2}$, \eqref{eq:x2s_gg_kt_approximation}, in addition to the standard GB approximations \eqref{gb_constrains_1} and \eqref{gb_constrains_2}.

\subsubsection*{Diagram 2}
The Feynman rules for diagram 2 give
\begin{align}
 i \mathcal{M}_2^{qq'\arr qq'g} &= \frac{F_{2}}{q^2 (p_{1}-k)^2} \, \Gamma_{ws}^{\mu\alpha}(p_{3},p_{1}-k,p_{1}) \breakeq &\qquad \times \epsilon_{\alpha}^{*}(k) \, g_{\mu\nu} \, J_{w's'}^{\nu}(p_{4},p_{2}) \breakeq
&\simeq \frac{F_{2}\,\delta_{ws} \delta_{w's'}}{q^2 (p_{1}-k)^2} \, (2p_1+q-2k)^\mu (2p_2-q)_\mu \breakeq &\qquad \times (2p_1 - k)^\alpha \epsilon_{\alpha}^{*}(k)
\end{align}
with $F_{2} = -i \, g^{3} \, \lambda^a_{jl} \lambda^c_{mn} \lambda^b_{ij} \delta^{ac}$. The kinematic approximations
\begin{align}
(p_{1}-k)^2 &\simeq -\frac{k_{\perp}^2}{x}\\
(2p_1+q-2k)^\mu (2p_2-q)_\mu &\simeq 2s\, (1-x) \label{eq:M2_x2kt_approx}\\
(2p_1 - k)^\alpha \epsilon_{\alpha}^{*} &\simeq \frac{2}{x} {\bf k}_{\perp}\! \cdot\! {\boldsymbol \epsilon}_\perp
\end{align}
together with $q^2 \simeq -q_{\perp}^{2}$ lead to the result listed in Eq.~\eqref{gb_me_single_diagrams}. Again Eq.~\eqref{eq:M2_x2kt_approx} requires the restriction to forward emission, $x^2 s \gg k_{\perp}^{2}$, in addition to the standard GB approximations \eqref{gb_constrains_1} and \eqref{gb_constrains_2}.

\subsubsection*{Diagram 3}
In order to explicitly compute diagrams 3 and 4 the components of the momentum transfer $q$ need to be redetermined from the on-shell conditions ($p_3^2=(p_1+q)^2=0$ and $p_4^2=(p_2-q-k)^2=0$) and the choice of keeping $k$ in the form given in Eq.~\eqref{eq:k_definition},
\begin{equation}
q^\mu \simeq \l( -  x\sqrs, \frac{q_\perp^{2}}{\sqrs}, {\bf q}_\perp  \r) \ . 
\end{equation}
With this definition diagram 3 can be evaluated as
\begin{align}
 i \mathcal{M}_3^{qq'\arr qq'g} &= \frac{F_{3}}{q^2 (p_{4}+k)^2} \, J_{ws}^{\mu}(p_{3},p_{1}) \, g_{\mu\nu} \breakeq &\qquad \times \Gamma_{w's'}^{\alpha\nu}(p_{4},p_{4}+k,p_{2}) \epsilon_{\alpha}^{*}(k) \breakeq
&\simeq \frac{F_{3}\,\delta_{ws} \delta_{w's'}}{q^2 (p_{2}-q)^2} \, (2p_1+q)^\mu (2p_2-q)_\mu \breakeq &\qquad \times (2p_2 - 2q-k)^\alpha \epsilon_{\alpha}^{*}(k)
\end{align}
with $F_{3} = -i \, g^{3} \, \lambda^a_{il} \lambda^c_{mj} \lambda^b_{jn} \delta^{ac}$. Employing the usual kinematic approximations simplifies the single terms
\begin{align}
(p_{2}-q)^2 &\simeq x s  \label{eq:M3_x2kt_approx}\\
(2p_1+q)^\mu (2p_2-q)_\mu &\simeq s\, (2-x)\\
(2p_2 -2q - k)^\alpha \epsilon_{\alpha}^{*} &\simeq 2 ( {\bf q}_{\perp} + {\bf k}_{\perp} )\! \cdot\! {\boldsymbol \epsilon}_\perp \ .
\end{align}
Combining these approximations yields
\begin{equation}
 i \mathcal{M}_3^{qq'\arr qq'g} \simeq F_{3}\,\delta_{ws} \delta_{w's'} \frac{2s}{q_{\perp}^{2}} \, \frac{2-x}{s}\, ( {\bf q}_{\perp} + {\bf k}_{\perp} ) \cdot {\boldsymbol \epsilon}_\perp
\end{equation}
which is suppressed by $1/s$ compared to the other diagrams listed in \eqref{gb_me_single_diagrams} and thus in the GB approximations justifies setting $i \mathcal{M}_3^{qq'\arr qq'g} \simeq 0$.

\subsubsection*{Diagram 4}
Diagram 4 is given by
\begin{align}
 i \mathcal{M}_4^{qq'\arr qq'g} &= \frac{F_{4}}{q^2 (p_{2}-k)^2} \, J_{ws}^{\mu}(p_{3},p_{1}) \, g_{\mu\nu} \breakeq &\qquad \times \Gamma_{w's'}^{\nu\alpha}(p_{4},p_{2}-k,p_{2}) \epsilon_{\alpha}^{*}(k) \breakeq
&\simeq \frac{F_{4}\,\delta_{ws} \delta_{w's'}}{q^2 (p_{2}-k)^2} \, (2p_1+q)^\mu (2p_2-q-2k)_\mu \breakeq &\qquad \times (2p_2-k)^\alpha \epsilon_{\alpha}^{*}(k)
\end{align}
with $F_{4} = -i \, g^{3} \, \lambda^a_{il} \lambda^c_{jn} \lambda^b_{mj} \delta^{ac}$. Since $(2p_2-k)^\alpha \epsilon_{\alpha}^{*} = 0$ without any further approximation, diagram 4 does not contribute.

\subsubsection*{Diagram 5}
Diagram 5 that describes the gluon emission off the exchanged gluon is the most tedious one to evaluate. It is given by
\begin{align}
 i \mathcal{M}_5^{qq'\arr qq'g} &= \frac{F_{5}}{q^2 (q-k)^2} \, J_{ws}^{\mu'}(p_{3},p_{1}) \, J_{w's'}^{\nu'}(p_{4},p_{2}) \, g_{\mu\mu'} g_{\nu\nu'} \breakeq 
&\qquad \times \epsilon_{\alpha}^{*}(k) \left[ g^{\mu\nu}(k-2q)^{\alpha} + g^{\nu\alpha}(q+k)^{\mu} \right. \breakeq &\qquad \quad \left. + g^{\alpha\mu}(q-2k)^{\nu} \right] \breakeq
&\simeq \frac{F_{5}\,\delta_{ws} \delta_{w's'}}{q^2 (q-k)^2} \, \left[ A + B + C \right]
\end{align}
with $F_{5} = g^{3} \, f^{abc} \lambda^a_{il} \lambda^c_{mn}$ and
\begin{align}
 A &= \left[ (2p_1+q-k)^{\mu}(q+k)_{\mu} \right]\,\left[ (2p_2-q)^{\alpha} \epsilon_{\alpha}^{*} \right] \breakeq
   &\simeq \frac{2-x}{1-x} \, \frac{k_\perp^2}{x} \, {\bf q}_{\perp}\!\cdot\! {\boldsymbol \epsilon}_\perp \\ 
 B &= \left[ (2p_2-q)^{\mu}(q-2k)_{\mu} \right]\,\left[ (2p_1+q-k)^{\alpha} \epsilon_{\alpha}^{*} \right] \breakeq
   &\simeq -4s \, {\bf k}_{\perp}\!\cdot\! {\boldsymbol \epsilon}_\perp \\ 
 C &= \left[ (2p_1+q-k)^{\mu}(2p_2-q)_{\mu} \right]\,\left[ (k-2q)^{\alpha} \epsilon_{\alpha}^{*} \right] \breakeq
  &\simeq 2s(2-x) \, {\bf q}_{\perp}\!\cdot\! {\boldsymbol \epsilon}_\perp \ .\label{eq:M5_x2kt_approx}
\end{align}
The term $A$ is not proportional to $s$ and thus can be neglected in respect to $B$ and $C$.
Together with the approximations for the propagators, $q^{2} \simeq -q_{\perp}^{2}$ and
\begin{equation}
(q-k)^2 \simeq -\frac{( {\bf q}_\perp - {\bf k}_\perp )^2}{(1-x)} \ ,
\end{equation}
this leads to the result in Eq.~\eqref{gb_me_single_diagrams}, where additionally approximation \eqref{gb_constrains_2} needs to be employed in collecting the above given terms. The constraint $x^2 s \gg k_{\perp}^{2}$ is needed in the computation of diagram 5 when approximating the first term of $C$ as given in \eqref{eq:M5_x2kt_approx}.

\bibliography{hq}

\end{document}